\documentclass{svjour3}                     % onecolumn (standard format)
\usepackage{graphicx}
\usepackage{fix-cm}
\usepackage{bm}
\usepackage[T1]{fontenc}
\usepackage{paralist}
\usepackage{amssymb,amsmath}
\usepackage{mathptmx}      % use Times fonts if available on your TeX system
\begin{document}

\title{The transformation of irreducible tensor operators under spherical functions
}

\titlerunning{The transformation of irreducible tensor operators}        % if too long for running head

\author{Rytis Jur\v{s}\.{e}nas         \and
        Gintaras Merkelis %etc.
}


\institute{R. Jur\v{s}\.{e}nas \at
              Institute of Theoretical Physics and Astronomy of Vilnius University, A. Go\v{s}tauto 12, LT-01108, Vilnius, 		              Lithuania \\
              Tel.: +370 5 2125361\\
              \email{Rytis.Jursenas@tfai.vu.lt}           %  \\
           \and
           G. Merkelis \at
              Institute of Theoretical Physics and Astronomy of Vilnius University, A. Go\v{s}tauto 12, LT-01108, Vilnius,      			  Lithuania \\
              Tel.: +370 5 2610502\\
              \email{Gintaras.Merkelis@tfai.vu.lt}
}

\date{Received: date / Accepted: date}

\maketitle

\begin{abstract}
The irreducible tensor operators and their tensor products employing Racah 
algebra are studied. Transformation procedure of the coordinate system operators act on are introduced. The rotation matrices and their parametrization 
by the spherical coordinates of vector in the fixed and rotated coordinate systems 
are determined. A new way of calculation of the irreducible 
coupled tensor product matrix elements is suggested. As an example, the proposed technique is applied for the matrix element construction for two electrons in a field of a fixed nucleus.
%Include keywords, PACS and mathematical
%subject classification numbers as needed.
\end{abstract}

\section{Introduction}
\label{intro}
The main aim of present work is to parametrize irreducible matrix representation of 
either $SO(3)$ or $SU(2)$ group by the coordinates of $S^{2}\times S^{2}$, where 
$S^{2}$ denotes the unit $2$-dimension sphere. The motivation is grounded on the following 
occasions: (i) the difficulties in theoretical atomic spectroscopy arising through multiple integrals of $N$-electron angular parts; (ii) inconvenience of Wigner-Eckart theorem application for irreducible tensor operator matrix elements on the basis of 
functions, expressed in terms of Wigner $D$-function.

In theoretical atomic physics the algorithms of matrix element
calculation for atomic quantities on
the basis of many-electron wave functions are well known and widely
used \cite{Condon,Jucys}. The construction of matrix element is based on the structure of many-electron function, which is represented by coupled tensor product of one-electron eigenstates. The latter formulation leads to complicated $N$-electron angular parts and various techniques, in order to simplify the calculation of many-electron matrix element \cite{Gaigalas,Grant,Abbadi}. According to Racah \cite{Racah,Racah2}, the basis eigenstates $\phi_{m}^{\lambda}$
behave in the same way as the spherical tensor operators $T_{m}^{\lambda}$
by means of transformations under irreducible matrix representations. The eigenstates $\phi_{m}^{\lambda}$ of central-field
atomic Hamiltonian are usually enunciated by spherical functions on $SO(3)/SO(2)$ quotient group, multiplied by $2\times 1$ spin matrices. However, in \cite{Rudzkas2} it was showed that $\phi\left(\hat{x}_{\xi}\right)=\phi\left(\theta_{\xi},\varphi_{\xi}\right)$ can be expressed by Wigner $D\left(\bar{\Omega}_{\xi}\right)$
functions on $SU(2)$, where $\bar{\Omega}_{\xi}=\left(\Phi_{\xi},\Theta_{\xi},0\right)=\left(\varphi_{\xi}+\pi /2,
\theta_{\xi},0\right)$. Furthermore, Bhatia \emph{et al.} in \cite{Bhatia} constructed two-electron wave function 
$\Phi\left(\hat{x}_{1},\hat{x}_{2}\right)$
in terms of the spherical functions $D\left(\Omega\right)$, where $\Omega$,
as usually, denotes Euler angles $(\Phi,\Theta,\Psi)$, i.e., rotation on
$S^{2}$ from $\hat{x}_{1}$ to $\hat{x}_{2}$
and \emph{vice verse}. Unfortunately, it is evident that for this basis the Wigner-Eckart theorem can not be applied directly, what leads to necessity to expand $D\left(\Omega\right)$ over $\hat{x}_{1},\hat{x}_{2}$.

In this work we start from
notations in \cite{Condon}, \cite{Jucys2}. We express rotation matrix by the spherical coordinates of vector in the fixed
and rotated coordinate systems, rather than the rotation angles in an
explicit form (Sec. \ref{geometry} and Sec. \ref{transform}). Following this route, we demonstrate technique of matrix element construction, when $2N$-integral (over the spherical
coordinates $\theta_{\xi},\varphi_{\xi}$ with $\xi=1,2,\ldots,N$) is reduced up to a double one (Sec. \ref{RCGC}). 
The technique is based on studied properties of integrity for obtained spherical functions (Sec. \ref{integr}) and proposed transformation coefficients, called the 
rotated Clebsch-Gordan coefficients (CGC) or simply RCGC (Sec. \ref{transform2}).

\section{\label{deff}Preliminaries}

The well known transformation formula for
$k$-rank spherical tensor operators $T_{q}^{k}$, $q\in \left[-k,+k\right]$, is given by

\begin{equation}
T_{q}^{k}\left(K_{2}\right)={\displaystyle \sum_{q^{\prime}}}D_{qq^{\prime}}^{k}
\left(\Omega\right)T_{q^{\prime}}^{k}\left(K_{1}\right),\label{eq:2.1}\end{equation}

\noindent{}where the symbols $K_1$ and $K_2$ show that the corresponding tensor operator is defined in the fixed and 
rotated coordinate system, respectively. The generalized spherical function $D_{qq^{\prime}}^{k}$ on $SO(3)$ or $SU(2)$
is of the form \cite{Jucys2}

\begin{equation} 
D_{qq^{\prime}}^{k}\left(\Omega\right)=a\left(k,q,q^{\prime}\right){\textstyle \mathrm{e}^{\mathrm{i}
\left(q\Phi+q'\Psi\right)}}\left\{ \cos\left({\scriptstyle \frac{1}{2}}\Theta\right)\right\}^{2k}
{\displaystyle\sum_{p}}b_{p}\left(k,q,q^{\prime}\right)\left\{ \tan\left({\scriptstyle \frac{1}{2}}\Theta\right)
\right\}^{2p-q+q^{\prime}},\label{eq:2.2}\end{equation}

\begin{equation}
a\left(k,q,q^{\prime}\right)=\mathrm{i}^{q'-q}\sqrt{\left(k+q\right)!\left(k-q\right)!\left(k+q'\right)!\left(k-q'\right)!},
\label{eq:2.3}\end{equation}

\begin{equation}
b_{p}\left(k,q,q^{\prime}\right)=\frac{(-1)^{p}}{p!\left(p+q'-q\right)!\left(k+q-p\right)!\left(k-q'-p\right)!},
\label{eq:2.4}\end{equation}

\noindent{}where $\Phi,\Psi\in\left[0,2\pi\right]$ and $\Theta\in\left[0,\pi\right]$.
In this work the standard phase system is used in which
complex and Hermitian conjugate operators accordingly denote $\overline{D_{qq^{\prime}}^{k}}=(-1)^{q-q^{\prime}}
D_{-q-q^{\prime}}^{k}$
and $T_{q}^{k\dagger}=(-1)^{k-q}T_{-q}^{k}$. It follows from Eq. (\ref{eq:2.1}), the rotated coordinate system $K_2$ depends on the fixed
system $K_1$ and the rotation angles $\Omega$. Such a dependency can be reformulated in a different way: 
the rotation angles $\Omega$ depend on the coordinates of given (or known) point, located on $S^{2}$ in the 
fixed and rotated coordinate systems. These spherical coordinates will be denoted $\hat{x}_{1}$ (in $K_{1}$) and 
$\hat{x}_{2}$ (in $K_2$). On the other hand, the spherical function $D$, which depends on $\Omega$,
can also be expressed as a function of the variables $\hat{x}_{1}$,
$\hat{x}_{2}$. In the next section the latter dependency
in an explicit form is presented.

\section{\label{geometry}The geometry of rotation angles}

In order to express the generalized spherical function
$D$ by the coordinates $\hat{x}_{1}$, $\hat{x}_{2}$
of vector in the fixed and rotated coordinate system, the
following geometry is defined. Suppose we have a map $\Omega:S^{2}\times S^{2}\mapsto SO\left(3\right)$.
Its representation onto the linear space of vectors $\hat{r}_{i}=\bm{r}_{i}/\left|\bm{r}_{i}\right|=\left(\begin{array}{ccc}
x_{i} & y_{i} & z_{i}\end{array}\right)^{\mathrm{T}}\in\mathbf{R}^{3}$ ($\left|\bm{r}_{i}\right|=\mathrm{const.}\:\forall i\in
\mathbb{Z}^{+}$ - positive integers)
is associated in the following way

\begin{equation}
\left\{ \begin{array}{l}
\hat{r}_{2}=R\left(\Omega\right)\hat{r}_{1},\\
x_{i}=\sin\theta_{i}\cos\varphi_{i}, y_{i}=\sin\theta_{i}\sin\varphi_{i}, z_{i}=\cos\theta_{i},\end{array}\right.
\label{eq:3.1}\end{equation}

\noindent{}where the coordinate system $K_1$ is rotated (the $ZXZ$ convention)
by the rotation matrix $R$ of $SO\left(3\right)$ \cite{Jucys2,Gelfand}. Then the system
of three equations in Eq. (\ref{eq:3.1}) is rewritten in the form

\begin{equation}
\left\{ \begin{array}{l}
x_{2}=-x_{1}^{\prime}\sin\varphi-u^{\prime}\cos\varphi,\\
y_{2}=x_{1}^{\prime}\cos\varphi-u^{\prime}\sin\varphi,\\
z_{2}=y_{1}^{\prime}\sin\theta+z_{1}^{\prime}\cos\theta,\end{array}\right.\label{eq:3.4}\end{equation}

\noindent{}where

\begin{equation}
 u^{\prime}=y_{1}^{\prime}\cos\theta-z_{1}^{\prime}\sin\theta,\label{eq:3.5a}
\end{equation}

\begin{equation}
\left\{ \begin{array}{l}
x_{1}^{\prime}=x_{1}\cos\Psi-y_{1}\sin\Psi,\\
y_{1}^{\prime}=x_{1}\sin\Psi+y_{1}\cos\Psi,\\
z_{1}^{\prime}=z_{1}.\end{array}\right.\label{eq:3.5}
\end{equation}

\noindent{}The parameter $\Psi$ is chosen optionally in the range
$\left[0,2\pi\right]$. Partial solutions of subsystems $(x_{2},z_{2})$ and $(y_{2},z_{2})$ for $\theta,\varphi$ are substituted then in Eq. (\ref{eq:2.2}). Optimal values of $\Psi$ are found by solving variational equation, varying obtained spherical function with respect to $\Psi$ (for details see Appendix \ref{A}). We attain, that solutions in $\mathbb{R}$ satisfy equation

\begin{equation}
\sin\theta_{1}\cos\left(\varphi_{1}+\Psi\right)=0.\label{eq:3.6.14}\end{equation}

\noindent{}The solutions are:
\begin{enumerate}
\item \label{i}$\cos\left(\varphi_{1}+\Psi\right)=0\Rightarrow\Psi=-\varphi_{1}+\sigma^{\prime}\frac{\pi}{2}+\bar{\sigma}\pi+2\pi n^{\prime},\: n^{\prime}\in\mathbb{Z}^{+},\: \sigma^{\prime}=\pm 1,\,\bar{\sigma}=0,\pm 1.$
\item \label{ii}$\sin\theta_{1}=0\quad\forall\Psi\in[0,2\pi].$
\end{enumerate}
\noindent{}First of all let us analyze item (\ref{i}) - the situation
when $\sin\theta_{1}\neq0$. Then $y_{1}^{\prime}=\sigma^{\prime}\sin\theta_{1}$, if $\bar{\sigma}=0$ and
$y_{1}^{\prime}=-\sigma^{\prime}\sin\theta_{1}$ if $\bar{\sigma}=\pm 1$
(see Eq. (\ref{eq:3.5})). Secondly, when item (\ref{ii})
is valid, we obtain that $\theta_{1}=0$ or $\theta_{1}=\pi$. The
parameter $\Psi$ then could be of any value in the range $\left[0,2\pi\right]$. 
It is clear that the solution $\sin\theta_{1}=0$ (item (\ref{ii}))
is a particular case of item (\ref{i}) if the angle $\Psi$ is
chosen to be equal to $\Psi=-\varphi_{1}+\sigma^{\prime}\frac{\pi}{2}+\bar{\sigma}\pi+2\pi n^{\prime}$. Since in
item (\ref{ii}) the solution $\Psi$ can have arbitrary values in the range $\left[0,2\pi\right]$, 
we choose it equal to the solution given by item (\ref{i}). Consequently, the solutions of the system in Eq. (\ref{eq:3.4}) are

\begin{equation}
\Phi=\varphi_{2}+\alpha\frac{\pi}{2},\:\Theta=\beta\left(\theta_{1}-\gamma\theta_{2}\right)+2\pi n,\:\Psi=-\varphi_{1}+\delta\frac{\pi}{2}+2\pi n^{\prime},\label{eq:g}\end{equation}

\noindent{}where $n^{\prime}\in\mathbb{Z}^{+}$ and the values for $\alpha$, $\beta$,
$\gamma$, $\delta$, $n$ are presented in Table \ref{tab:Tab1}. The function $\Omega(\hat{x}_{1},\hat{x}_{2})$ may be expanded into several different geometries, representing miscellaneous rotations.

\begin{table}
\caption{\label{tab:Tab1}The values for the parameters $\alpha$, $\beta$,
$\gamma$, $\delta$, $n$}
\begin{tabular}{|c|c|c|c|c|c|c|c||c|c|c|c|c|c|c|}
\hline 
\multicolumn{1}{|c}{The maps} & \multicolumn{1}{c}{} &  & $\alpha$ & $\beta$ & $\gamma$ & $\delta$ & $n$ & \multicolumn{1}{c}{The maps} &  & $\alpha$ & $\beta$ & $\gamma$ & $\delta$ & $n$\tabularnewline
\hline
\hline 
 & $\Omega_{1}^{+}$ & $\Omega_{11}^{+}$ & $+$ & $+$ & $+$ & $-$ &  &  & $\Omega_{2}^{+}$ & $+$ & $+$ & $-$ & $+$ & $0$\tabularnewline
\cline{3-7} 
$\Omega_{1}^{\pm}$ &  & $\Omega_{12}^{+}$ & $-$ & $+$ & $+$ & $+$ & $0$ & $\Omega_{2}^{\pm}$ &  &  &  &  &  & \tabularnewline
\cline{2-7} \cline{10-15} 
 & $\Omega_{1}^{-}$ & $\Omega_{11}^{-}$ & $+$ & $-$ & $+$ & $-$ &  &  & $\Omega_{2}^{-}$ & $-$ & $-$ & $-$ & $-$ & $1$\tabularnewline
\cline{3-7} 
 &  & $\Omega_{12}^{-}$ & $-$ & $-$ & $+$ & $+$ &  &  &  &  &  &  &  & \tabularnewline
\hline
\end{tabular}
\end{table}

\begin{inparaenum}[(a)]
\item $\theta_{1}-\theta_{2}\in\left[0,\pi\right]$.
\end{inparaenum}

\begin{equation}
\begin{array}{c}
\varphi_{2}\in\left[0,\pi\right]:\\
\\\Omega_{11}^{+}=\left\{ \begin{array}{ll}
\Phi=\varphi_{2}+\frac{\pi}{2},\\
\Theta=\theta_{1}-\theta_{2},\\
\Psi=-\varphi_{1}-\frac{\pi}{2} & \left(2\pi\right).\end{array}\right.\end{array}\qquad\begin{array}{c}
\varphi_{2}\in\left(\pi,2\pi\right]:\\
\\\Omega_{12}^{+}=\left\{ \begin{array}{ll}
\Phi=\varphi_{2}-\frac{\pi}{2},\\
\Theta=\theta_{1}-\theta_{2},\\
\Psi=-\varphi_{1}+\frac{\pi}{2} & \left(2\pi\right).\end{array}\right.\end{array}\label{eq:o1}\end{equation}

\begin{inparaenum}[(b)]
\item $\theta_{2}-\theta_{1}\in\left[0,\pi\right]$.
\end{inparaenum}

\begin{equation}
\begin{array}{c}
\varphi_{2}\in\left[0,\pi\right]:\\
\\\Omega_{11}^{-}=\left\{ \begin{array}{ll}
\Phi=\varphi_{2}+\frac{\pi}{2},\\
\Theta=\theta_{2}-\theta_{1},\\
\Psi=-\varphi_{1}-\frac{\pi}{2} & \left(2\pi\right).\end{array}\right.\end{array}\qquad\begin{array}{c}
\varphi_{2}\in\left(\pi,2\pi\right]:\\
\\\Omega_{12}^{-}=\left\{ \begin{array}{ll}
\Phi=\varphi_{2}-\frac{\pi}{2},\\
\Theta=\theta_{2}-\theta_{1},\\
\Psi=-\varphi_{1}+\frac{\pi}{2} & \left(2\pi\right).\end{array}\right.\end{array}\label{eq:o2}\end{equation}

\begin{inparaenum}[(c)]
\item $\theta_{1}+\theta_{2}\in\left(0,\pi\right]$.
\end{inparaenum}

\begin{equation}
\begin{array}{c}
\Omega_{2}^{+}=\left\{ \begin{array}{ll}
\Phi=\varphi_{2}+\frac{\pi}{2},\\
\Theta=\theta_{1}+\theta_{2},\\
\Psi=-\varphi_{1}+\frac{\pi}{2} & \left(2\pi\right).\end{array}\right.\end{array}\label{eq:o3}\end{equation}

\begin{inparaenum}[(d)]
\item $\theta_{1}+\theta_{2}\in\left[\pi,2\pi\right]$.
\end{inparaenum}

\begin{equation}
\begin{array}{c}
\Omega_{2}^{-}=\left\{ \begin{array}{ll}
\Phi=\varphi_{2}-\frac{\pi}{2},\\
\Theta=2\pi-\theta_{1}-\theta_{2},\\
\Psi=-\varphi_{1}-\frac{\pi}{2} & \left(2\pi\right).\end{array}\right.\end{array}\label{eq:o4}\end{equation}

\noindent{}The function $\Omega_{1}^{+}=\left\{ \Omega_{11}^{+},\Omega_{12}^{+}\right\} $
is matched for the case when $\theta_{1}\geq\theta_{2}$, while
the function $\Omega_{1}^{-}=\left\{ \Omega_{11}^{-},\Omega_{12}^{-}\right\} $
is matched for the case when $\theta_{2}\geq\theta_{1}$. The functions
$\Omega_{11}^{\pm}$ describe rotations for given $\varphi_{2}\in\left[0,\pi\right]$;
according to $\Omega_{12}^{\pm}$, rotations are realized for
$\varphi_{2}\in\left(\pi,2\pi\right]$. The function $\Omega_{2}^{\pm}=\left\{ \Omega_{2}^{+},\Omega_{2}^{-}\right\} $
defines another possible rotation for the given angles $\theta_{1}+\theta_{2}\in\left(0,\pi\right]$
or $\theta_{1}+\theta_{2}\in\left[\pi,2\pi\right]$. Note, if $\varphi_{2}\in\left[\frac{3\pi}{2},2\pi\right]$,
then for the rotation $\Omega_{2}^{+}$, the angle
$\Phi>2\pi$. On the other hand, rotation over the angle
$2\pi$ geometrically is equivalent to initial state.
Thus, one can choose whether $\Phi>2\pi$ or $0<\Phi-2\pi<2\pi$.
The same idea is valid and for the rotation $\Omega_{2}^{-}$. Finally,
when $\theta_{1}=\theta_{2}=0$, the angle $\Theta=0$ and the angles
$\Phi$, $\Psi$ acquire any values in $\left[0,2\pi\right]$.
Then the rotation matrix $R\left(\Omega\right)=R_{z}\left(\Phi+\Psi\right)$.
Hence, in this case the full rotation is made by the angle $\Phi+\Psi$
around the $z$-axis. In other words, if $\theta_{1}=\theta_{2}=0$,
rotation by the Euler angles $\Omega$ is not singularly defined.
Further it will be assumed that $\theta_{1}$ and $\theta_{2}$
are not equal to zero at the same time (that is why the range of $\theta_{1}+\theta_{2}$ for
$\Omega_{2}^{+}$ is open from the left).

\section{\label{transform}Spherical functions}

In the previous section the mapping from $S^{2}\times S^{2}$ to $SO\left(3\right)$ has been defined. It was demonstrated that possible rotations in $\mathbf{R}^{3}$ from $K_{1}$ to $K_{2}$ could be realized by the rotation angles $\Omega_{1}^{\pm}=\left\{\Omega_{1}^{+},\Omega_{1}^{-}\right\}=\left\{\Omega_{11}^{+},\Omega_{12}^{+},\Omega_{11}^{-},\Omega_{12}^{-}\right\}$ and $\Omega_{2}^{\pm}=\left\{\Omega_{2}^{+},\Omega_{2}^{-}\right\}$. If substituting these functions in Eq. (\ref{eq:2.2}), we would obtain the following spherical function (for alternative expressions, see Appendix \ref{B})

\[
\left(n,n^{\prime};\alpha,\beta,\gamma,\delta|\hat{x}_{1},\hat{x}_{2}\right)_{qq^{\prime}}^{k}=\mathrm{i}^{\alpha q+\delta q^{\prime}}(-1)^{2\left(nk+n^{\prime}q^{\prime}\right)}\beta^{q^{\prime}-q}a\left(k,q,q^{\prime}\right)\mathrm{e}^{\mathrm{i}\left(q\varphi_{2}-q^{\prime}\varphi_{1}\right)}\]

\begin{equation}
\times\left\{ \cos\left[{\scriptstyle \frac{1}{2}}\left(\theta_{1}-\gamma\theta_{2}\right)\right]\right\} ^{2k}{\displaystyle \sum_{p}}b_{p}\left(k,q,q^{\prime}\right)\left\{ \tan\left[{\scriptstyle \frac{1}{2}}\left(\theta_{1}-\gamma\theta_{2}\right)\right]\right\} ^{2p+q^{\prime}-q},\label{eq:4}\end{equation}

\noindent{}where $k\in\mathbb{Z}^{+},\mathbb{Q}^{+}$ and $\mathbb{Q}^{+}=\left\{m+1/2;m\in\mathbb{Z}^{+}\right\}$; the indices $q,q^{\prime}\in\left[-k,+k\right]$. Let us define particular cases of Eq. (\ref{eq:4}) in the
following way

\begin{equation}
\left(0,n^{\prime};+,\pm,+,-|\hat{x}_{1},\hat{x}_{2}\right)_{qq^{\prime}}^{k}=
\;^{\pm}\xi_{qq^{\prime}}^{k}\left(\hat{x}_{1},\hat{x}_{2}\right),\label{eq:14}\end{equation}

\begin{equation}
\left(0,n^{\prime};-,\pm,+,+|\hat{x}_{1},\hat{x}_{2}\right)_{qq^{\prime}}^{k}=\;^{\pm}\vartheta_{qq^{\prime}}^{k}\left(\hat{x}_{1},\hat{x}_{2}\right)=(-1)^{q^{\prime}-q}\:\;^{\pm}\xi_{qq^{\prime}}^{k}\left(\hat{x}_{1},\hat{x}_{2}\right),\label{eq:15}\end{equation}

\begin{equation}
\left(0,n^{\prime};+,+,-,+|\hat{x}_{1},\hat{x}_{2}\right)_{qq^{\prime}}^{k}=\;^{+}\zeta_{qq^{\prime}}^{k}\left(\hat{x}_{1},\hat{x}_{2}\right),\label{eq:12}\end{equation}

\begin{equation}
\left(1,n^{\prime};-,-,-,-|\hat{x}_{1},\hat{x}_{2}\right)_{qq^{\prime}}^{k}=\;^{-}\zeta_{qq^{\prime}}^{k}\left(\hat{x}_{1},\hat{x}_{2}\right)=(-1)^{2q^{\prime}}\:\;^{+}\zeta_{qq^{\prime}}^{k}\left(\hat{x}_{1},\hat{x}_{2}
\right),\label{eq:13}\end{equation}

\noindent{}and the matrices

\begin{equation}
\begin{array}{cl}
\Omega_{1}^{\pm}: & \;^{\pm}\eta^{k}\left(\hat{x}_{1},\hat{x}_{2}\right)\in\left\{ \;^{\pm}\xi^{k}\left(\hat{x}_{1},\hat{x}_{2}\right),\;^{\pm}\vartheta^{k}\left(\hat{x}_{1},\hat{x}_{2}\right)\right\}.\end{array}\label{eq:16}\end{equation}

\noindent{}It is seen, the spherical functions $^{\pm}\eta_{qq^{\prime}}^{k}\left(\hat{x}_{1},\hat{x}_{2}\right)$
and $^{\pm}\zeta_{qq^{\prime}}^{k}\left(\hat{x}_{1},\hat{x}_{2}\right)$
are the generalized spherical functions $D_{qq^{\prime}}^{k}\left(\Omega_{1}^{\pm}\right)$
and $D_{qq^{\prime}}^{k}\left(\Omega_{2}^{\pm}\right)$ parametrized by the coordinates of
$S^{2}\times S^{2}$, respectively. Particularly, the matrix $^{+}\eta^{k}\left(\hat{x}_{1},\hat{x}_{2}\right)$
represents rotation in $\mathbf{R}^{3}$ from $K_{1}$ to $K_{2}$, when $\theta_{1}\geq\theta_{2}$
and it is associated to the map $\Omega_{1}^{+}$, while the matrix
$^{-}\eta^{k}\left(\hat{x}_{1},\hat{x}_{2}\right)$
describes rotation from $K_{1}$ to $K_{2}$ when $\theta_{2}\geq\theta_{1}$
and it is associated to the map $\Omega_{1}^{-}$. The matrices $^{+}\zeta^{k}\left(\hat{x}_{1},\hat{x}_{2}\right)$ and $^{-}\zeta^{k}\left(\hat{x}_{1},\hat{x}_{2}\right)$ are related to the maps $\Omega_{2}^{+}$ and $\Omega_{2}^{-}$.

In accordance with Eqs. (\ref{eq:14})-(\ref{eq:13}), the spherical functions $^{\pm}\xi$, $^{\pm}\vartheta$ and $^{\pm}\zeta$ are connected to each other as follows

\begin{equation}
\begin{array}{lcl}
\;^{-}\xi_{qq^{\prime}}^{k}=(-1)^{q^{\prime}-q}\:\;^{+}\xi_{qq^{\prime}}^{k}, &  & \;^{-}\vartheta_{qq^{\prime}}^{k}=(-1)^{q^{\prime}-q}\:\;^{+}\vartheta_{qq^{\prime}}^{k},\\
\;^{+}\vartheta_{qq^{\prime}}^{k}=\;^{-}\xi_{qq^{\prime}}^{k}, &  & \;^{-}\zeta_{qq^{\prime}}^{k}=(-1)^{2q^{\prime}}\:\;^{+}\zeta_{qq^{\prime}}^{k}.\end{array}\label{eq:rel}\end{equation}

\noindent{}It directly follows from Eq. (\ref{eq:rel}), that

\begin{equation}
\;^{+}\eta_{qq^{\prime}}^{k}=\;^{-}\eta_{qq^{\prime}}^{k}=\eta_{qq^{\prime}}^{k}\in\left\{ \;^{+}\xi_{qq^{\prime}}^{k},\;^{-}\xi_{qq^{\prime}}^{k}\right\} ,\quad\;^{-}\xi_{qq^{\prime}}^{k}=\overline{\;^{+}\xi_{-q-q^{\prime}}^{k}},\label{eq:relA}\end{equation}

\noindent{}where the column vectors of $^{\pm}\xi^{k}$ are orthonormal, i.e.,

\begin{equation}
{\displaystyle \sum_{q}}\;^{+}\xi_{qq^{\prime}}^{k}\:\;^{-}\xi_{-q-q^{\prime\prime}}^{k}=\delta_{q^{\prime}q^{\prime\prime}}.\label{eq:relB}\end{equation}

\noindent{}The latter condition is, of course, valid and for the rest of spherical functions.

It is noticeable,
the matrices $\eta^{k}$ and $^{\pm}\zeta^{k}$ are the unitary irreducible
matrix representations of $SO\left(3\right)$ (for $k\in\mathbb{Z}^{+}$) or of $SU\left(2\right)$ (for $k\in\mathbb{Q}^{+}$), parametrized by the
coordinates of $S^{2}\times S^{2}$. Hence, the irreducible
tensor operators $T_{q}^{k}$ transform
among themselves as follows (to compare, see Eq. (\ref{eq:2.1}))

\begin{equation}
\begin{array}{lc}
\Omega_{1}^{\pm}: & T_{q}^{k}\left(K_{2}\right)={\displaystyle \sum_{q^{\prime}}}\eta_{qq^{\prime}}^{k}\left(\hat{x}_{1},\hat{x}_{2}\right)T_{q^{\prime}}^{k}\left(K_{1}\right),\\
\Omega_{2}^{\pm}: & T_{q}^{k}\left(K_{2}\right)={\displaystyle \sum_{q^{\prime}}}\;^{\pm}\zeta_{qq^{\prime}}^{k}\left(\hat{x}_{1},\hat{x}_{2}\right)T_{q^{\prime}}^{k}\left(K_{1}\right).\end{array}\label{eq:tr17}\end{equation}

\noindent{}The transformation formula for the maps $\Omega_{1}^{\pm}$ is
restricted by the condition $\theta_{1}\neq\theta_{2}$.
In a contrary case, only the maps $\Omega_{2}^{\pm}$ are valid. Consequently, the reduction formulas for the spherical functions 

\begin{equation}
\;^{\pm}\tau_{qq^{\prime}}^{k}\left(.,.\right)\in\left\{ \eta_{qq^{\prime}}^{k}\left(.,.\right),\;^{\pm}\zeta_{qq^{\prime}}^{k}\left(.,.\right)\right\} \label{eq:18a}\end{equation}

\noindent{}are these

\begin{equation}
\;^{\pm}\tau_{q_{1}q_{1}^{\prime}}^{k_{1}}\left(.,.\right)\:\;^{\pm}\tau_{q_{2}q_{2}^{\prime}}^{k_{2}}\left(.,.\right)={\displaystyle \sum_{k}}\;^{\pm}\tau_{qq^{\prime}}^{k}\left(.,.\right)\left[\begin{array}{ccc}
k_{1} & k_{2} & k\\
q_{1} & q_{2} & q\end{array}\right]\left[\begin{array}{ccc}
k_{1} & k_{2} & k\\
q_{1}^{\prime} & q_{2}^{\prime} & q^{\prime}\end{array}\right],\label{eq:18}\end{equation}

\noindent{}where in the brackets $\left(.,.\right)$ the spherical
coordinates of $S^{2}\times S^{2}$ are given. The
Clebsch-Gordan coefficients of $SU\left(2\right)$ are none zero only
when $q=q_{1}+q_{2}$ and $q^{\prime}=q_{1}^{\prime}+q_{2}^{\prime}$.
The summation is performed over $k=\left|k_{1}-k_{2}\right|,\left|k_{1}-k_{2}\right|+1,\ldots,k_{1}+k_{2}$.

\paragraph*{Example.}\label{ex2}

$\:$ Suppose, $\hat{x}_{1}=\left(\frac{\pi}{6},\frac{\pi}{4}\right)$
and $\hat{x}_{2}=\left(\frac{\pi}{3},\pi\right)$. Possible rotations are realized then in accordance with
$\Omega_{11}^{-}=\left(\frac{3\pi}{2},\frac{\pi}{6},\frac{5\pi}{4}\right)$, $\Omega_{2}^{+}=\left(\frac{3\pi}{2},\frac{\pi}{2},\frac{\pi}{4}\right)$.
Transformation formulas in Eq. (\ref{eq:tr17}) are valid for the spherical
functions $^{-}\xi_{qq^{\prime}}^{k}\left(\frac{\pi}{6},\frac{\pi}{4},\frac{\pi}{3},\pi\right)$
and $^{+}\zeta_{qq^{\prime}}^{k}\left(\frac{\pi}{6},\frac{\pi}{4},\frac{\pi}{3},\pi\right)$
which coequal to $D_{qq^{\prime}}^{k}\left(\frac{3\pi}{2},\frac{\pi}{6},\frac{5\pi}{4}\right)$
and $D_{qq^{\prime}}^{k}\left(\frac{3\pi}{2},\frac{\pi}{2},\frac{\pi}{4}\right)$.
Suppose $k=\frac{5}{2}$ and $q=-\frac{1}{2}$, $q^{\prime}=\frac{3}{2}$.
Then

\[
\;^{-}\xi_{-\frac{1}{2}\frac{3}{2}}^{\frac{5}{2}}\left(\textstyle\frac{\pi}{6},\frac{\pi}{4},\frac{\pi}{3},\pi\right)=D_{-\frac{1}{2}\frac{3}{2}}^{\frac{5}{2}}\left(\textstyle\frac{3\pi}{2},\frac{\pi}{6},\frac{5\pi}{4}\right)=\textstyle\frac{(-1)^{\frac{1}{8}}}{32}\left(13-3\sqrt{3}\right),\]

\[
\;^{+}\zeta_{-\frac{1}{2}\frac{3}{2}}^{\frac{5}{2}}\left(\textstyle\frac{\pi}{6},\frac{\pi}{4},\frac{\pi}{3},\pi\right)=D_{-\frac{1}{2}\frac{3}{2}}^{\frac{5}{2}}\left(\textstyle\frac{3\pi}{2},\frac{\pi}{2},\frac{\pi}{4}\right)=\textstyle\frac{(-1)^{\frac{5}{8}}}{4}.\]

\noindent{}One may notice, according to expressions in Eq. (\ref{eq:rel}), it is possible to find out how all other spherical functions are related to the calculated functions above.

\section{\label{integr}The integral of spherical functions}

The choice of geometries $\Omega_{1}^{\pm}$, $\Omega_{2}^{\pm}$
is convenient for other applications of obtained spherical functions
$^{\pm}\tau_{qq^{\prime}}^{k}$. This is because the parameters $\theta_{1},\theta_{2}$
and $\varphi_{1},\varphi_{2}$ are separated. Consequently, the functions
which depend on these parameters can be integrated separately, i.e.,
the integral on $S^{2}$ is naturally separated into the
integrals over $\varphi_{i}$ and over $\theta_{i}$. 
By definition, the functions $^{\pm}\tau_{qq^{\prime}}^{k}\left(\hat{x}_{1},\hat{x}_{2}\right)$ are determined 
in different areas $L^{2}(\Omega)\subset S^{2}$, which restrict the existence of integrity. Therefore, we construct the integral

\begin{equation}
\mathcal{S}_{qq^{\prime}}^{k}\left(\hat{x}_{1};\gamma\right)={\displaystyle \int_{S^{2}}}\mathrm{d}\hat{x}_{2}\:\left(n,n^{\prime};\alpha,\beta,\gamma,\delta|\hat{x}_{1},\hat{x}_{2}\right)_{qq^{\prime}}^{k}.
\label{eq:33s000}\end{equation}

Suppose, $\gamma=+1$, i.e., the map $\Omega_{1}^{\pm}$ is realized. Let the areas (or paths of integration) on $S^{2}$ to be

\begin{equation}
\begin{array}{lc}
L^{2}\left(\Omega_{11}^{+}\right)=\left\{ \varphi_{2}\in\left[0,\pi\right];\theta_{2}\in\left[0,\theta_{1}\right]\right\} , & L^{2}\left(\Omega_{12}^{+}\right)=\left\{ \varphi_{2}\in\left[\pi,2\pi\right];\theta_{2}\in\left[0,\theta_{1}\right]\right\} ,\\
L^{2}\left(\Omega_{11}^{-}\right)=\left\{ \varphi_{2}\in\left[0,\pi\right];\theta_{2}\in\left[\theta_{1},\pi\right]\right\} , & L^{2}\left(\Omega_{12}^{-}\right)=\left\{ \varphi_{2}\in\left[\pi,2\pi\right];\theta_{2}\in\left[\theta_{1},\pi\right]\right\} .\end{array}\label{eq:42}\end{equation}

\noindent{}Here $\Omega_{11}^{\pm}$ and $\Omega_{12}^{\pm}$
mark the existence of integrable
spherical functions $^{\pm}\xi_{qq^{\prime}}^{k}\left(\hat{x}_{1},\hat{x}_{2}\right)$
in the corresponding areas $L^{2}\left(\Omega_{11}^{\pm}\right)$ and $L^{2}\left(\Omega_{12}^{\pm}\right)$. The parameters $\alpha$, $\beta$, $\gamma$, $\delta$, $n$
are determined then according to the values presented in Table \ref{tab:Tab1} and the equations
in Eq. (\ref{eq:g}). It is clear, the spherical function $^{+}\xi$ is integrable in $L^{2}\left(\Omega_{11}^{+}\right)$ with $n^{\prime}\in\left\{1,2\right\}$ and in $L^{2}\left(\Omega_{12}^{-}\right)$ with $n^{\prime}\in\left\{0,1\right\}$; the function $^{-}\xi$ is integrable in $L^{2}\left(\Omega_{11}^{-}\right)$ with $n^{\prime}\in\left\{1,2\right\}$ and in $L^{2}\left(\Omega_{12}^{+}\right)$ with $n^{\prime}\in\left\{0,1\right\}$. But $^{-}\xi_{qq^{\prime}}^{k}=(-1)^{q^{\prime}-q}\:\;^{+}\xi_{qq^{\prime}}^{k}$ (see Eq. (\ref{eq:rel})). This implies

$\begin{array}{ll}

\mathcal{S}_{qq^{\prime}}^{k}\left(\hat{x}_{1};+\right)&={\displaystyle \int_{L^{2}\left(\Omega_{11}^{+}\right)}}\mathrm{d}\hat{x}_{2}\:\;^{+}\xi_{qq^{\prime}}^{k}\left(\hat{x}_{1},\hat{x}_{2}\right)+{\displaystyle \int_{L^{2}\left(\Omega_{11}^{-}\right)}}\mathrm{d}\hat{x}_{2}\:\;^{-}\xi_{qq^{\prime}}^{k}\left(\hat{x}_{1},\hat{x}_{2}\right)\\

&+{\displaystyle \int_{L^{2}\left(\Omega_{12}^{+}\right)}}\mathrm{d}\hat{x}_{2}\:\;^{-}\xi_{qq^{\prime}}^{k}\left(\hat{x}_{1},\hat{x}_{2}\right)+{\displaystyle \int_{L^{2}\left(\Omega_{12}^{-}\right)}}\mathrm{d}\hat{x}_{2}\:\;^{+}\xi_{qq^{\prime}}^{k}\left(\hat{x}_{1},\hat{x}_{2}\right)\\

&=\lambda_{q^{\prime}}\left(\varphi_{1}\right)\mathrm{i}^{q-q^{\prime}-1}{\textstyle \frac{(-1)^{q}-1}{q}}\left((-1)^{q^{\prime}}+1\right)a\left(k,q,q^{\prime}\right)\mathrm{e}^{-\mathrm{i}q^{\prime}\varphi_{1}}
\end{array}$

\begin{equation}
\times{\displaystyle \sum_{p}}b_{p}\left(k,q,q^{\prime}\right)\left(\:_{p}I_{qq^{\prime}}^{k}\left(\theta_{1};+;0,\theta_{1}\right)+(-1)^{q-q^{\prime}}\:_{p}I_{qq^{\prime}}^{k}\left(\theta_{1};+;\theta_{1},\pi\right)\right),\label{eq:S2c}\end{equation}

\begin{equation}
\lambda_{q^{\prime}}\left(\varphi_{1}\right)=\left\{ \begin{array}{ll}
(-1)^{q^{\prime}}, & \quad\varphi_{1}\in\left[0,\frac{\pi}{2}\right],\\
(-1)^{2q^{\prime}}, & \quad\varphi_{1}\in\left(\frac{\pi}{2},\frac{3\pi}{2}\right],\\
(-1)^{3q^{\prime}}, & \quad\varphi_{1}\in\left(\frac{3\pi}{2},2\pi\right].\end{array}\right.\label{eq:S2c1}\end{equation}

\noindent{}The definition of $\,_{p}I_{qq^{\prime}}^{k}$ is given by the formula

\begin{equation}
\,_{p}I_{qq^{\prime}}^{k}\left(\theta_{1};\gamma ; a,b\right)={\displaystyle \int_{a}^{b}}\mathrm{d}\theta_{2}\:\sin\theta_{2}\:\left\{ \cos\left[{\scriptstyle \frac{1}{2}}\left(\theta_{1}-\gamma\theta_{2}\right)\right]\right\} ^{2k}\left\{ \tan\left[{\scriptstyle \frac{1}{2}}\left(\theta_{1}-\gamma\theta_{2}\right)\right]\right\} ^{2p+q^{\prime}-q}.\label{eq:22}\end{equation}

\noindent{}For $\gamma=+1$ the integration
is performed making the change of integrand $z=\tan\left[\left(\theta_{1}-\theta_{2}\right)/2\right]$.
After some ordinary trigonometric manipulations it acquires the form

\begin{equation}
\,_{p}I_{qq^{\prime}}^{k}\left(\theta_{1};+;a,b\right)=2\left\{ 2I_{1}^{p+}\left(a,b\right)\cos\theta_{1}+\left(I_{2}^{p+}\left(a,b\right)-I_{0}^{p+}\left(a,b\right)\right)\sin\theta_{1}\right\} ,\label{eq:36I}\end{equation}

\begin{equation}
I_{s}^{p+}\left(a,b\right)=I_{s}^{p+}\left(\tan\frac{\theta_{1}-b}{2}\right)-I_{s}^{p+}\left(\tan\frac{\theta_{1}-a}{2}\right)\label{eq:37I}\end{equation}

\noindent{}with $s=0,1,2$ and $I_{s}^{p+}\left(z\right)$ defined by

\[
I_{s}^{p+}\left(z\right)={\displaystyle \int_{\mathbb{R}}}\mathrm{d}z\:\frac{z^{2p+q^{\prime}-q+s}}
{\left(1+z^{2}\right)^{k+2}}=\frac{z^{2p+q^{\prime}-q+s+1}}{2p+q^{\prime}-q+s+1}\]

\begin{equation}
\times\:\,_{2}F_{1}\left(\frac{2p+q^{\prime}-q+s+1}{2},k+2;\frac{2p+q^{\prime}-q+s+3}{2};-z^{2}\right)+\textrm{const.},\label{eq:25}\end{equation}

\noindent{}where $\,_{2}F_{1}$ denotes Gauss hypergeometric function.

When $\theta_{1}=0$ or $\theta_{1}=\pi$, the function $I_{s}^{p+}\left(0,\pi\right)$
depends on infinite variables ($z=\pm\infty$). For this reason bellow
the boundary values of $\,_{p}I_{qq^{\prime}}^{k}\left(\theta;+;0,\pi\right)$
are presented.

\begin{enumerate}
\item $\theta=0$. Then $\,_{p}I_{qq^{\prime}}^{k}\left(0;+;0,\pi\right)=4I_{1}^{p+}\left(-\infty\right)$
(since $I_{1}^{p+}\left(0\right)=0$), where\\
\begin{equation}
I_{1}^{p+}\left(-\infty\right)=\frac{(-1)^{q^{\prime}-q}}{2}B\left(k+1-p+\frac{q-q^{\prime}}{2},1+p+\frac{q^{\prime}-q}{2}\right)\label{eq:30}\end{equation}
\\
with $B$ being Beta function.
\item $\theta=\pi$. Then\\
\begin{equation}
\,_{p}I_{qq^{\prime}}^{k}\left(\pi;+;0,\pi\right)=(-1)^{q-q^{\prime}}\,_{p}I_{qq^{\prime}}^{k}\left(0;+;0,\pi\right).\label{eq:31}\end{equation}

\end{enumerate}
For example, it directly follows from Eqs. (\ref{eq:22}), (\ref{eq:31}), that

\begin{equation}
I_{2}={\displaystyle \int_{0}^{\pi}}\mathrm{d}\theta\:\sin^{2k+1}\theta\:\cos^{\gamma}\theta=\left[1+\left(-1\right)^{
\gamma}\right]I_{1}^{p}\left(\infty\right),\label{eq:cite}\end{equation}

\begin{equation}
I_{1}^{p}\left(\infty\right)=\frac{1}{2}B\left(k+1,\frac{\gamma+1}{2}\right),\label{eq:cite2}\end{equation}

\noindent{}where the left hand side of $I_{2}$ is the same integral discussed by Pinchon \emph{et al.} \cite{Pinchon}
(Eqs. (19)-(20)), when developing rotation matrices for real spherical
harmonics.

When $k\in\mathbb{Z}^{+}$, the integral of the spherical function $\eta$ acquires the form

\[
\mathcal{S}_{qq^{\prime}}^{k}\left(\hat{x}_{1};+\right)=\delta_{q0}\pi\mathrm{i}^{-q^{\prime}}\left((-1)^{q^{\prime}}+1\right)a\left(k,0,q^{\prime}\right)\mathrm{e}^{-\mathrm{i}
q^{\prime}\varphi_{1}}
\]

\begin{equation}
\times{\displaystyle \sum_{p}}b_{p}\left(k,0,q^{\prime}\right)\:_{p}I_{0q^{\prime}}^{k}\left(\theta_{1};+;0,\pi\right).\label{eq:S2d}\end{equation}

\noindent{}Hence, for integer $k$, the integral of $\left(0,n^{\prime};\alpha,\beta,+,\delta|\hat{x}_{1},\hat{x}_{2}\right)_{qq^{\prime}}^{k}$ is none zero only if $q=0$ and $q^{\prime}$ is even. Particularly, $\mathcal{S}_{00}^{0}\left(\hat{x}_{1};+\right)=4\pi$.

Finally, the integration of $\left(n,n^{\prime};\alpha,\beta,-,\delta|\hat{x}_{1},\hat{x}_{2}\right)_{qq^{\prime}}^{k}$ must
be proceeded. For the maps $\Omega_{2}^{\pm}$ there is no difference which
value of $\theta_{1}$ and $\theta_{2}$ is greater or less (or equal).
It has been demonstrated earlier, that for
$\Omega_{2}^{+}$ the condition $0\leq\theta_{1}+\theta_{2}\leq\pi$
($\alpha=\beta=\delta=+$, $n=0$) must be satisfied, while
for $\Omega_{2}^{-}$ the restriction is: $\pi\leq\theta_{1}+\theta_{2}\leq2\pi$
($\alpha=\beta=\delta=-$, $n=1$). But when integrating
$^{\pm}\zeta_{qq^{\prime}}^{k}\left(\hat{x}_{1},\hat{x}_{2}\right)$
over $\theta_{2}$, the angle $\theta_{2}$ acquires all values in
$\left[0,\pi\right]$. This means the map $\Omega_{2}^{+}$
is valid only if $\theta_{1}=0$ and the map $\Omega_{2}^{-}$ is
realized only for $\theta_{1}=\pi$. Hence, the integration of $^{\pm}\zeta_{qq^{\prime}}^{k}\left(\hat{x}_{1},\hat{x}_{2}\right)$
can not be correctly performed for any values of $\theta_{1}$, except
for $\theta_{1}=0$ or $\theta_{1}=\pi$. In other words, the spherical functions $^{\pm}\zeta_{qq^{\prime}}^{k}\left(\hat{x}_{1},\hat{x}_{2}\right)$, which represent rotations $\Omega_{2}^{\pm}$, are not integrable on $S^{2}$, in general. For this reason we conclude,
that the most preferable spherical functions (at least for integration)
are $\eta_{qq^{\prime}}^{k}\left(\hat{x}_{1},\hat{x}_{2}\right)$,
i.e., those which represent geometries $\Omega_{1}^{\pm}$. Note, the angles $\left(\theta_{1},\varphi_{1}\right)$ and $\left(\theta_{2},\varphi_{2}\right)$ are fully determined on $S^{2}$ for the maps $\Omega_{1}^{\pm}$. This implies that for any values of $\theta_{1}$, $\varphi_{1}$, $\theta_{2}$, $\varphi_{2}$ there will always exist at least one rotation from the set $\Omega_{1}^{\pm}=\left\{\Omega_{11}^{\pm},\Omega_{12}^{\pm}\right\}$. Consequently, the loss of geometry $\Omega_{2}^{\pm}$ does not imply the loss of generality by means of the existence of at least one spherical function from the set $\left\{^{+}\xi,\,^{-}\xi\right\}$.

\section{\label{RCGC}RCGC technique}

Since the irreducible matrix representations $\eta^{\lambda}\left(\hat{x}_{1},\hat{x}_{2}\right)$
depend on known coordinates $\hat{x}_{1}$ and $\hat{x}_{2}$,
it is worth to exploit them in the study of tensor products
of irreducible tensor operators $T$ (or basis functions $\phi$) directly, but not formally, as in most cases,
when tensor operators are transformed under representations
$D^{\lambda}\left(\Omega\right)$. Moreover, most of the physical operators $T_{q}^{k}$, basically studied in atomic
spectroscopy, are expressed in terms of $D$ and their various combinations. These are, for example,
the spherical operators $C_{q}^{k}(\hat{x})=\mathrm{i}^{k}D_{q0}^{k}(\bar{\Omega})$, the spherical harmonics 
$Y_{q}^{k}(\hat{x})=\sqrt{\left(2k+1\right)/4\pi}C_{q}^{k}(\hat{x})$. Expressions over $D$ 
of other operators, such as spin operator $S^{1}$, angular momentum operator $L^{1}$, can be found, 
for instance, in \cite{Rudzkas2}. All these mentioned operators also transform among themselves according to Eq. (\ref{eq:tr17}). 
Hence, it is natural for such operators
to write $T(\hat{x})$ instead of $T(K)$, which is a general case. We say the operator 
$T(\hat{x})$ acts on $\hat{x}$ coordinate. Below the latter notation will be used.

\subsection{\label{transform2}Transformation coefficients}

Reduction formula for tensor product reads (see Eq. (\ref{eq:tr17}))

\begin{equation}
 T_{m_{1}}^{\lambda_{1}}\left(\hat{x}_{1}\right)T_{m_{2}}^{\lambda_{2}}\left(\hat{x}_{2}\right)=
{\displaystyle \sum_{\lambda}}\bar{T}_{m}^{\lambda}\left(\hat{x}_{1}\right)
 c_{m_{1}m_{2}m}^{\lambda_{1}\lambda_{2}\lambda}\left(\hat{
x}_{1},\hat{x}_{2}\right),\label{eq:rr5.1}\end{equation}

\noindent{}In this paper the so-called for simplicity rotated
Clebsch-Gordan coefficient of the first type or simply RCGC I is defined by

\begin{equation}
c_{m_{1}m_{2}m}^{\lambda_{1}\lambda_{2}\lambda}\left(\hat{x}_{1},\hat{x}_{2}\right)=
{\displaystyle \sum_{m_{2}^{\prime}}}\eta_{m_{2}m_{2}^{\prime}}^{\lambda_{2}}\left(\hat{x}_{1},
\hat{x}_{2}\right)\left[\begin{array}{ccc}
\lambda_{1} & \lambda_{2} & \lambda\\
m_{1} & m_{2}^{\prime} & m\end{array}\right],\label{eq:rr5.2}\end{equation}

\noindent{}From Eq. (\ref{eq:rr5.1}) we obtain the next expression

\begin{equation}
 \bar{\bar{T}}_{m}^{\lambda}\left( \hat{x}_{1},\hat{x}_{2}\right)={\displaystyle \sum_{\lambda^{\prime}m^{\prime}}} C_{m^{\prime}m}^{\lambda_{1}\lambda_{2}
\lambda^{\prime}\lambda}\left(\hat{x}_{1},\hat{x}_{2}\right)
 \bar{T}_{m^{\prime}}^{\lambda^{\prime}}\left(\hat{x}_{1}\right),\label{eq:rr5.3}\end{equation}

\noindent{}where rotated Clebsch-Gordan coefficient of the second
type or simply RCGC II is

\begin{equation}
 C_{m^{\prime}m}^{\lambda_{1}\lambda_{2}\lambda^{\prime}\lambda}\left(\hat{x}_{1},\hat{x}_{2}\right)=
{\displaystyle \sum_{m_{1}m_{2}}}c_{m_{1}m_{2}m^{\prime}}^{\lambda_{1}\lambda_{2}\lambda^{\prime}}\left(\hat{x}_{1},
\hat{x}_{2}\right)
\left[\begin{array}{ccc}
\lambda_{1} & \lambda_{2} & \lambda\\
m_{1} & m_{2} & m\end{array}\right].\label{eq:rr5.4}\end{equation}

\noindent{}Irreducible tensor operators $\bar{T}_{m}^{\lambda}\left(\hat{x}_{1}\right)$ and $\bar{\bar{T}}_{m}^{\lambda}\left( \hat{x}_{1},\hat{x}_{2}\right)$ are delivered by applying reduction rules for the Kronecker product $\lambda_{1}\times\lambda_{2}\rightarrow \lambda$. However, $\bar{T}_{m}^{\lambda}$ acts on $\hat{x}_{1}$, while $\bar{\bar{T}}_{m}^{\lambda}$ acts on $\hat{x}_{1}$, $\hat{x}_{2}$. For example, if $T$ represents normalized spherical harmonic $C$, then coupled tensor product $\bar{T}_{m}^{\lambda}\left(\hat{x}_{1}\right)=\mathrm{i}^{\lambda_{1}+\lambda_{2}-\lambda}C_{m}^{\lambda}
\left(\hat{x}_{1}\right)\Bigl[\begin{smallmatrix}\lambda_{1}& \lambda_{2}& \lambda\\ 0& 0 &0\end{smallmatrix}\Bigr]$. Contrarily, tensor operator $\bar{\bar{T}}_{m}^{\lambda}\left( \hat{x}_{1},\hat{x}_{2}\right)$ can not be reduced into one $C_{m}^{\lambda}$ due to different spaces, in which $T_{m_{i}}^{\lambda_{i}}\left(\hat{x}_{i}\right)$ ($i=1,2$) act on.

Another useful circumstance for RCGC application is based on a possibility to reduce these coefficients. In accordance with Eq. (\ref{eq:18}), we directly attain

\[
c_{m_{1}m_{2}m}^{\lambda_{1}\lambda_{2}\lambda}(.,.)c_{\bar{m}_{1}\bar{m}_{2}\bar{m}}^{\bar{\lambda}_{1}\bar{\lambda}_{2}\bar{\lambda}}(.,.)={\displaystyle \sum_{\Lambda_{2}}}\eta_{M_{2}M_{2}^{\prime}}^{\Lambda_{2}}(.,.){\displaystyle \sum_{m_{2}^{\prime}\bar{m}_{2}^{\prime}}}\left[\begin{array}{ccc}
\lambda_{1} & \lambda_{2} & \lambda\\
m_{1} & m_{2}^{\prime} & m\end{array}\right]\left[\begin{array}{ccc}
\bar{\lambda}_{1} & \bar{\lambda}_{2} & \bar{\lambda}\\
\bar{m}_{1} & \bar{m}_{2}^{\prime} & \bar{m}\end{array}\right]\]

\begin{equation}
\times\left[\begin{array}{ccc}
\lambda_{2} & \bar{\lambda}_{2} & \Lambda_{2}\\
m_{2} & \bar{m}_{2} & M_{2}\end{array}\right]\left[\begin{array}{ccc}
\lambda_{2} & \bar{\lambda}_{2} & \Lambda_{2}\\
m_{2}^{\prime} & \bar{m}_{2}^{\prime} & M_{2}^{\prime}\end{array}\right].\label{eq:ccred}\end{equation}

\noindent{}Combining Eqs. (\ref{eq:rr5.4})-(\ref{eq:ccred}), we may also derive reduction formula for RCGC II. It is obvious, demonstrated reduction procedure (see Eq. (\ref{eq:rr5.3})) can be extended and for many-electron wave functions. 

Suppose, for instance, one needs to calculate matrix element of $T_{q}^{k}\left(\hat{x}\right)$ on the basis of eigenfunctions $Y_{m}^{l}\left(\hat{x}\right)$, described on $SO(3)/SO(2)$. The application of RCGC technique directly indicates, that in present case

\begin{equation}
[l\Vert T^{k}\Vert\bar{l}]=\mathrm{i}^{\bar{l}-l}[(2\bar{l}+1)/(2l+1)]^{1/2}T_{0}^{k}(\hat{0})\left[\begin{array}{ccc}
\bar{l} & k & l\\
0 & 0 & 0\end{array}\right].\label{eq:ccred2}\end{equation}

\noindent{}The proof is produced in Appendix \ref{C}. Here it is assumed $\hat{0}\equiv (0,0)$. Reduced matrix element $[l\Vert T^{k}\Vert\bar{l}]$ is obtained from Wigner-Eckart theorem $\langle lm\vert T_{q}^{k}\vert \bar{l}\bar{m}\rangle=(-1)^{2k}[l\Vert T^{k}\Vert\bar{l}]\Bigl[\begin{smallmatrix}\bar{l}& k& l\\ \bar{m}& q &m\end{smallmatrix}\Bigr]$. Particularly, if $T^{k}=C^{k}$, then operator $T_{0}^{k}\left(\hat{0}\right)=\mathrm{i}^{k}$, and obtained formula is in fully consistence with Eq. (2.52) in \cite{Jucys}. Thus, Eq. (\ref{eq:ccred2}) is a generalization of this special case. Going on this route, one can derive matrix element expressions on the basis of more complex eigenfunctions. 

One can see from Eq. (\ref{eq:rr5.3}), the specific feature of technique, based on coordinate
transformations (or simply RCGC technique), is that
tensor structure of operator and wave functions is preserved,
and the resultant matrix element is calculated on the basis
of transformed operators $\bar{T}\left(\hat{x}\right)$,
i.e., the calculation of multiple integrals transforms
to the calculation of a single integral (over $\hat{x}$). This is because each $N$-electron matrix element

$$\begin{array}{l}
{\displaystyle \int_{S^{2}}}\mathrm{d}\hat{x}_{1}{\displaystyle \int_{S^{2}}}
\mathrm{d}\hat{x}_{2}\ldots{\displaystyle \int_{S^{2}}}\mathrm{d}\hat{
x}_{N}\:\Phi_{M}^{\Lambda^{bra}\dagger}\left(\hat{x}_{1},\hat{x}_{2},\ldots,
\hat{x}_{N}\right)\\

\times T_{Q}^{K}\left(\hat{x}_{1},\hat{x}_{2},\ldots,\hat{x}_{N}
\right)\Phi_{M^{\prime}}^{\Lambda^{ket}}\left(\hat{x}_{1},\hat{x}_{2},\ldots,
\hat{x}_{N}\right)
\end{array}$$

\noindent{}is transformed to

$$\begin{array}{l}
{\displaystyle \int_{S^{2}}}\mathrm{d}\hat{x}_{1}\:\bar{\Phi}_{m}^{\lambda^{bra}\dagger}
\left(\hat{x}_{1}\right)\bar{T}_{q}^{k}\left(\hat{x}_{1}\right)\bar{
\Phi}_{m^{\prime}}^{\lambda^{ket}}\left(\hat{x}_{1}\right)\\

\times{\displaystyle \int_{S^{2}}}\mathrm{d}\hat{x}_{2}{\displaystyle \int_{S^{2}}}
\mathrm{d}\hat{x}_{3}\ldots{\displaystyle \int_{S^{2}}}\mathrm{d}\hat{
x}_{N}\:\overline{\eta_{m_{2}\mu_{2}}^{\lambda_{2}^{bra}}\left(\hat{x}_{1},
\hat{x}_{2}\right)}\:\:\:\overline{\eta_{m_{3}\mu_{3}}^{\lambda_{3}^{bra}}\left(\hat{x}_{1},
\hat{x}_{3}\right)}\ldots\overline{\eta_{m_{N}\mu_{N}}^{\lambda_{N}^{bra}}\left(\hat{x}_{1},
\hat{x}_{N}\right)}\\

\times\eta_{q_{2}\pi_{2}}^{k_{2}}\left(\hat{x}_{1},\hat{x}_{2}\right)
\eta_{q_{3}\pi_{3}}^{k_{3}}\left(\hat{x}_{1},\hat{x}_{3}\right)
\ldots\eta_{q_{N}\pi_{N}}^{k_{N}}\left(\hat{x}_{1},\hat{x}_{N}\right)\\

\times\eta_{m_{2}^{\prime}\mu_{2}^{\prime}}^{\lambda_{2}^{ket}}\left(\hat{x}_{1},\hat{x}_{2}\right)\eta_{m_{3}^{\prime}\mu_{3}^{\prime}}^{\lambda_{3}^{ket}}\left(\hat{x}_{1},\hat{x}_{3}\right)\ldots\eta_{m_{N}^{\prime}\mu_{N}^{\prime}}^{\lambda_{N}^{ket}}\left(\hat{x}_{1},\hat{x}_{N}\right),
\end{array}$$

\noindent{}where $\bar{\Phi}$ and $\bar{T}$ indicate coupled tensor products of $\phi^{\lambda_{i}}\left(\hat{x}_{1}\right)$ and $T^{k_{i}}\left(\hat{x}_{1}\right)$, respectively. The application of Eq. (\ref{eq:18}) for all $\overline{\eta_{m_{\xi}\mu_{\xi}}^{\lambda_{\xi}^{bra}}\left(\hat{x}_{1},\hat{x}_{\xi}\right)}$, $\eta_{q_{\xi}\pi_{\xi}}^{k_{\xi}}\left(\hat{x}_{1},\hat{x}_{\xi}\right)$, $\eta_{m_{\xi}^{\prime}\mu_{\xi}^{\prime}}^{\lambda_{\xi}^{ket}}\left(\hat{x}_{1},\hat{x}_{\xi}\right)$, $\xi=2,3,\ldots,N$ implies, that initially determined $2N$-integral reduces to a double one

$$
 {\displaystyle \int_{S^{2}}}\mathrm{d}\hat{x}_{1}\:\bar{\Phi}_{m}^{\lambda^{bra}\dagger}
\left(\hat{x}_{1}\right)\bar{T}_{q}^{k}\left(\hat{x}_{1}\right)
\mathcal{S}_{M_{2}M_{2}^{\prime}}^{\Lambda_{2}}\left(\hat{x}_{1};+\right)
\mathcal{S}_{M_{3}M_{3}^{
\prime}}^{\Lambda_{3}}\left(\hat{x}_{1};+\right)\ldots \mathcal{S}_{M_{N}M_{N}^{\prime}}^{\Lambda_{N}}
\left(\hat{x}_{1};+\right)\bar{\Phi}_{m^{\prime}}^{\lambda^{ket}}\left(\hat{x}_{1}\right).$$

\noindent{}Instead of that, for given $N$-electron wave functions we
produce $N-1$ functions $\mathcal{S}$ (see Eq. (\ref{eq:S2c}))
and a product of momenta coupling coefficients (CGC),
which can be decomposed into
$3nj$-coefficients, if summing over projections.

\subsection{\label{matrix2}Example}

A simple application of RCGC technique can be demonstrated, for example, in a study of two electrons,  
located in some external field (of fixed nucleus, for instance). We refer to \cite{Bhatia}, where a two-electron wave function is presented by

\begin{equation}
\Psi_{m}^{l}\left(\bm{r}_{1},\bm{r}_{2}\right)={\displaystyle \sum_{\mu}}g_{\mu}^{l}\left(r_{1},r_{2},
\theta_{12}\right)D_{m\mu}^{l}\left(\Omega\right),\label{eq:D10}\end{equation}

\noindent{}with $r_{i}=\left|\bm{r}_{i}\right|$; $l\in\mathbb{Z}^{+}$. 
In the same paper (Bhatia \emph{et al.} \cite{Bhatia}) it was determined, that Laplacian, involving
$\theta_{12}$ (the angle between vectors $\bm{r}_{1}$ and $\bm{r}_{2}$), does not affect the orbital angular momentum $l>0$. 
Thus we may assume, that $g$ is a radial function not
going into a deeper analysis, since our aim is the angular part. 
Then $\Psi_{m}^{l}$ may be rewritten as follows

\begin{equation}
\Psi_{m}^{l}\left(\bm{r}_{1},\bm{r}_{2}\right)={\displaystyle \sum_{\mu}}g_{\mu}^{l}\left(r_{1},r_{2},
\theta_{12}\right)\eta_{m\mu}^{l}\left(\hat{x}_{1},\hat{x}_{2}\right).
\label{eq:D12}\end{equation}

\noindent{}Suppose we want to calculate the Coulomb $1/r_{12}$ matrix element. In this case the interaction in a tensor form reads $T_{0}^{0}\left(\bm{r}_{1},\bm{r}_{2}\right)= 
\sum_{k} (r_{<}^{k}/r_{>}^{k+1})
\left(C^{k}\left(\hat{x}_{1}
\right)\cdot C^{k}\left(\hat{x}_{2}\right)\right)$, where $r_{<}=\mathrm{min}\left(r_{1},r_{2}\right)$ and
$r_{>}=\mathrm{max}\left(r_{1},r_{2}\right)$. The scalar product $\left(C^{k}\left(\hat{x}_{1}
\right)\cdot C^{k}\left(\hat{x}_{2}\right)\right)$ is reduced in agreement with
Eq. (\ref{eq:rr5.3}), what leads to the following construction of  
$\langle \Psi_{m}^{l} \vert r_{12}^{-1}\vert \Psi_{m^{\prime}}^{l^{\prime}}\rangle$ (taking into account only angular part)

$\begin{array}{ll}
\langle \Psi_{m}^{l} \vert r_{12}^{-1}\vert \Psi_{m^{\prime}}^{l^{\prime}}\rangle&=
(-1)^{m-m^{\prime}}{\displaystyle \sum_{\mu}}(-1)^{\mu}\overline{g_{\mu}^{l}\left(r_{1},r_{2},\theta_{12}\right)}g_{\mu+m-m^{\prime}}^{l^{\prime}}\left(r_{1},r_{2},\theta_{12}\right)\\

&\times{\displaystyle \sum_{k}} \frac{r_{<}^{k}}{r_{>}^{k+1}}
{\displaystyle \sum_{K}}{\displaystyle \sum_{Q=\mathrm{even}}}\mathrm{i}^{-K}\left[\begin{array}{ccc}
k & k & K\\
0 & 0 & 0\end{array}\right]
{\displaystyle \sum_{\bar{L}}}{\displaystyle \int_{S^{2}}}\mathrm{d}\hat{x}\: 
C_{Q}^{K}\left(\hat{x}\right)
\mathcal{S}_{0Q}^{\bar{L}}\left(\hat{x};+\right)\\

&\times{\displaystyle \sum_{L}}\left[\begin{array}{ccc}
k & k & K\\
m-m^{\prime} & Q+m^{\prime}-m & Q\end{array}\right]\left[\begin{array}{ccc}
l & k & L\\
-m & m-m^{\prime} & -m^{\prime}\end{array}\right]
\left[\begin{array}{ccc}
L & l^{\prime} & \bar{L}\\
-m^{\prime} & m^{\prime} & 0\end{array}\right]
\end{array}$

\begin{equation}
 \times\left[\begin{array}{ccc}
l & k & L\\
-\mu & Q+m^{\prime}-m & Q+m^{\prime}-m-\mu \end{array}\right]
\left[\begin{array}{ccc}
L & l^{\prime} & \bar{L}\\
Q+m^{\prime}-m-\mu & m-m^{\prime}+\mu & Q\end{array}\right].\label{eq:D17}\end{equation}

\noindent{}It is evident, this way of calculation is more efficient in comparison with direct integration
of $\iint_{S^{2}}\mathrm{d}\hat{x}_{1}\mathrm{d}
\hat{x}_{2}\: \overline{D_{m\mu}^{l}\left(\Omega\right)}\left(C^{k}\left(\hat{x}_{1}
\right)\cdot C^{k}\left(\hat{x}_{2}\right)\right)D_{m^{\prime}\mu^{\prime}}^{l^{\prime}}\left(\Omega\right)$, because otherwise one should change integrands, when mapping from $\hat{x}$
to $\Omega$. Consequently, that would lead to complex manipulations of trigonometric
equations, given in Eq. (\ref{eq:3.4}).

\section{\label{conclusions}Summary}

In present study, we produced a parametrization of standard Wigner $D$-function on $SU(2)$ by the coordinates $(\hat{x}_{1},\hat{x}_{2})$ of vector in the fixed and rotated coordinate systems. As a result, we found the set of spherical functions, conformed to $D(\Omega)$ in miscellaneous areas $L^{2}(\Omega)\subset S^{2}$. We showed that offered parametrization of $D(\Omega)$ provides an opportunity to reduce angular $2N$-integrals, which are of special interest in theoretical atomic spectroscopy, to a double one. Particularly, we demonstrated a new way of construction of irreducible tensor operator matrix element, which plays the role of a generalization of previously obtained special cases (see, for example, Eq. (\ref{eq:ccred2})). Another convenient usage of suggested RCGC technique, based on coordinate transformation, is applied to the calculation of matrix elements on the basis of functions, expressed in terms of standard $D(\Omega)$ functions (for two-electron case, see Sec. \ref{matrix2}).

\begin{acknowledgements}
The authors are grateful to R. Kisielius for attentive revision of this paper and for various useful remarks.
\end{acknowledgements}

\appendix

\section{\label{A}Optimal values for $\Psi$}

Here the solutions $\Omega=(\varphi+\pi/2,\theta,\Psi)\in\mathbb{R}\:\forall \hat{x}_{i}\in S^{2}$ of Eq. (\ref{eq:3.4}) will be studied in a more detail. Partial solutions of the systems $(x_{2},z_{2})$ and $(y_{2},z_{2})$ for $\theta,\varphi$ are

\begin{equation}
\theta_{\sigma_{1}\sigma_{2}}=\sigma_{1}\arccos\frac{z_{1}z_{2}+\sigma_{2}\left|y_{1}^{\prime}\right|\sqrt{z_{1}^{2}-z_{2}^{2}+y_{1}^{\prime2}}}{z_{1}^{2}+y_{1}^{\prime2}}+2\pi n_{1},\:n_{1}\in\mathbb{Z^{+}}, \label{eq:newsol1}\end{equation}

\begin{equation}
\varphi_{\sigma_{3}\sigma_{4}}^{(1)}=\sigma_{3}\arccos\frac{x_{1}^{\prime}y_{2}+\sigma_{4}\left|x_{2}\right|\sqrt{1-
x_{1}^{\prime2}-z_{2}^{2}}}{1-z_{2}^{2}}+2\pi n_{2}, \:n_{2}\in\mathbb{Z^{+}}, \label{eq:newsol2}\end{equation}

\begin{equation}
\varphi_{\sigma_{3}\sigma_{4}}^{(2)}=\sigma_{3}\arccos\frac{x_{2}\sqrt{1-x_{1}^{\prime2}-z_{2}^{2}}+\sigma_{4}\left|x_{1}^{\prime}y_{2}\right|}{\mathrm{sgn}\left(y_{1}^{\prime}\right)\left(1-z_{2}^{2}\right)}+2\pi n_{2},
\: n_{2}\in\mathbb{Z^{+}},
\label{eq:newsol3}\end{equation}

\noindent{}where $\sigma_{i}\in\{-1,+1\}$; $\mathrm{sgn}\left(y_{1}^{\prime}\right)$ denotes the sign of $y_{1}^{\prime}$. Let us mark the common values of $\varphi_{\sigma_{3}\sigma_{4}}^{(1)}$ and $\varphi_{\sigma_{3}\sigma_{4}}^{(2)}$ by $\varphi_{\sigma_{3}\sigma_{4}}=\,_{n_{2}}{\mathbf{w}}_{\sigma_{3}\sigma_{4}}\left(\hat{x}_{1},\hat{x}_{2};\Psi\right)$. Let us also denote

\begin{equation}
\sin^{2}\frac{\theta_{\sigma_{1}\sigma_{2}}}{2}=\mathbf{x}_{\sigma_{2}}\left(\hat{x}_{1},\hat{x}_{2};\Psi\right)
=\frac{z_{1}^{2}-z_{1}z_{2}+y_{1}^{\prime2}+\sigma_{2}y_{1}^{\prime}\sqrt{z_{1}^{2}-z_{2}^{2}+y_{1}^{\prime2}}}{2
\left(z_{1}^{2}+y_{1}^{\prime2}\right)}.
\label{eq:x}\end{equation}

\noindent{}Obtained functions $\mathbf{x}$ and $\mathbf{w}$ are substituted in Eq. (\ref{eq:2.2}). The generalized spherical function $D$ is rearranged to the following function

\[ \left\{ \,_{n_{2}}Z_{qq^{\prime}}^{k}\left(\hat{x}_{1},\hat{x}_{2};\Psi\right)\right\} _{
\sigma_{2}\sigma_{3}\sigma_{4}}=a\left(k,q,q^{\prime}\right)
\mathrm{e}^{\mathrm{i}q\left(\frac{\pi}{2}+\,_{n_{2}}\mathbf{w}_{
\sigma_{3}\sigma_{4}}\left(\hat{x}_{1},\hat{x}_{2};\Psi\right)\right)+\mathrm{i}q^{\prime}\Psi}{
\displaystyle \sum_{p}}b_{p}\left(k,q,q^{\prime}\right)\]

\begin{equation}
\times\mathbf{x}_{\sigma_{2}}^{p-\frac{q-q'}{2}}\left(\hat{x}_{1},\hat{x}_{2};\Psi\right)\left(1-\mathbf{x}_{
\sigma_{2}}\left(\hat{x}_{1},\hat{x}_{2};\Psi\right)\right)^{-p+k+\frac{q-q'}{2}}.\label{eq:3.7}\end{equation}

\noindent{}In order to look for optimal $\Psi$ values, we carry out
variational procedure for the gauge $\Psi$, applying
it for all possible distributions of $\sigma_{2}$, $\sigma_{3}$, $\sigma_{4}$, i.e., $\left(\delta/\delta\Psi\right)\,_{n_{2}}Z_{qq^{\prime}}^{k}=0$. This implies

\[
\mathrm{i}\,_{n_{2}}Z_{qq^{\prime}}^{k}\left(q^{\prime}+q\frac{\delta\,_{n_{2}}\mathbf{w}_{\sigma_{3}\sigma_{4}}}{\delta\Psi}\right)+a\mathrm{e}^{\mathrm{i}q\left(\frac{\pi}{2}+\,_{n_{2}}\mathbf{w}_{\sigma_{3}\sigma_{4}}\right)+\mathrm{i}q^{\prime}\Psi}{\displaystyle \sum_{p}}b_{p}\frac{\delta\mathbf{x}_{\sigma_{2}}}{\delta\Psi}\]

\begin{equation}
\times\mathbf{x}_{\sigma_{2}}^{p-\frac{q-q^{\prime}}{2}-1}\left(1-\mathbf{x}_{\sigma_{2}}\right)^{-p+k+\frac{q-q^{\prime}}{2}-1}\left(p-\frac{q-q^{\prime}}{2}-k\mathbf{x}_{\sigma_{2}}\right)=0.\label{eq:var1}\end{equation}

\noindent{}Applying obvious fact, that $A=B=0$ if $A,B\in\mathbb{R}$ in $A+\mathrm{i}B=0$, we finally gain

\begin{equation}
\left\{ \begin{array}{l}
\frac{\delta\mathbf{x}_{\sigma_{2}}}{\delta\Psi}=0,\\
\mathbf{x}_{\sigma_{2}}=\eta,\,\eta\in\{ 0,1,\frac{q^{\prime}-q}{2k}\} ,\\
q^{\prime}+q\frac{\delta\,_{n_{2}}\mathbf{w}_{\sigma_{3}\sigma_{4}}}{\delta\Psi}=0,\,q\neq 0.\end{array}\right.\label{eq:var2}\end{equation}

\noindent{}It follows from equation $\mathbf{x}_{\sigma_{2}}=\eta$,
that $y_{1}^{\prime}\in\mathbb{C}$. Thus, $\Psi$ does not belong to $\mathbb{R}$. Equation $\frac{\delta\mathbf{x}_{\sigma_{2}}}{\delta\Psi}=0$ is rewritten
in the form

\begin{equation}
\frac{\delta\mathbf{x}_{\sigma_{2}}}{\delta\Psi}=\frac{\delta\mathbf{x}_{\sigma_{2}}}{\delta y_{1}^{\prime}}\frac{\delta y_{1}^{\prime}}{\delta\Psi}=\frac{\delta\mathbf{x}_{\sigma_{2}}}{\delta y_{1}^{\prime}}x_{1}^{\prime}=0.
\label{eq:3.6.13}\end{equation}

\noindent{}The solutions of $\frac{\delta\mathbf{x}_{\pm}}{\delta y_{1}^{\prime}}=0$
for $y_{1}^{\prime}$ do not belong to $\mathbb{R}$. Thus, equation $x_{1}^{\prime}=0$ has to be solved. The latter
is equivalent to Eq. (\ref{eq:3.6.14}). Finally, when studying the third equation in Eq. (\ref{eq:var2}), we would
get some individual values of $\Psi$, which depend on $q,q^{\prime}$ (for $q=0$, present equation vanishes).
Thus, the set of solutions would be the subset of solutions given
by the first equation, independent of $q,q^{\prime}$.

\section{\label{B}The alternative expressions of spherical functions}

In accordance with Eqs. (\ref{eq:2.3})-(\ref{eq:2.4}), the product of coefficients $a$ and $b_{p}$ can be rewritten as follows

\begin{equation}
a\left(k,q,q^{\prime}\right)b_{p}\left(k,q,q^{\prime}\right)=\mathrm{i}^{q^{\prime}-q}(-1)^{p}\left[\frac{\left(k+q\right)!\left(k-q\right)!}{\left(k+q^{\prime}\right)!\left(k-q^{\prime}\right)!}\right]^{\frac{1}{2}}\left(\begin{array}{c}
k-q^{\prime}\\
p\end{array}\right)\left(\begin{array}{c}
k+q^{\prime}\\
p+q^{\prime}-q\end{array}\right),\label{eq:A01}\end{equation}

\noindent{}where the last two quantities on the right hand side of Eq. (\ref{eq:A01}) denote binomial coefficients. Further, let us mark the term

\begin{equation}
A_{qq^{\prime}}^{k}\left(\gamma\right)=a\left(k,q,q^{\prime}\right)\left\{ \cos\left[{\scriptstyle \frac{1}{2}}\left(\theta_{1}-\gamma\theta_{2}\right)\right]\right\} ^{2k}{\displaystyle \sum_{p}}b_{p}\left(k,q,q^{\prime}\right)\left\{ \tan\left[{\scriptstyle \frac{1}{2}}\left(\theta_{1}-\gamma\theta_{2}\right)\right]\right\} ^{2p+q^{\prime}-q}.\label{eq:A02}\end{equation}

\noindent{}According to Eq. (\ref{eq:A01}), $A_{qq^{\prime}}^{k}\left(\gamma\right)$ may be revised by

\begin{equation}
A_{qq^{\prime}}^{k}\left(\gamma\right)=\mathrm{i}^{q^{\prime}-q}\left[\frac{\left(k+q\right)!\left(k-q\right)!}{\left(k+q^{\prime}\right)!\left(k-q^{\prime}\right)!}\right]^{\frac{1}{2}}{\displaystyle \sum_{p}}(-1)^{p}\left(\begin{array}{c}
k-q^{\prime}\\
p\end{array}\right)\left(\begin{array}{c}
k+q^{\prime}\\
p+q^{\prime}-q\end{array}\right)\frac{z^{p+\frac{q^{\prime}-q}{2}}}{\left(1+z\right)^{k}},\label{eq:A03}\end{equation}

\noindent{}where $z=\tan^{2}\left[{\scriptstyle\frac{1}{2}}\left(\theta_{1}-\gamma\theta_{2}\right)\right]$. Performing the summation over $p$, we obtain

$$
{\displaystyle \sum_{p}}(-1)^{p}\left(\begin{array}{c}
k-q^{\prime}\\
p\end{array}\right)\left(\begin{array}{c}
k+q^{\prime}\\
p+q^{\prime}-q\end{array}\right)\frac{z^{p+\frac{q^{\prime}-q}{2}}}{\left(1+z\right)^{k}}$$

\begin{subequations}

\begin{equation}
=(-1)^{q-q^{\prime}}\left(\begin{array}{c}
k-q^{\prime}\\
q-q^{\prime}\end{array}\right) z^{\frac{q-q^{\prime}}{2}}\left(1+z\right)^{1+k}\:_{2}F_{1}\left(k+1+q,k+1-q^{\prime};1+q-q^{\prime};-z\right)\quad\left(q\geq q^{\prime}\right)
\label{eq:A04}\end{equation}

\begin{equation}
=\left(\begin{array}{c}
k+q^{\prime}\\
q^{\prime}-q\end{array}\right)z^{\frac{q^{\prime}-q}{2}}\left(1+z\right)^{1+k}\:_{2}F_{1}\left(k+1+q^{\prime},k+1-q;1+q^{\prime}-q;-z\right)\quad\left(q^{\prime}\geq q\right).\label{eq:A05}\end{equation}
\end{subequations}

\noindent{}Thus if: (a) $A_{qq^{\prime}}^{>k}=A_{qq^{\prime}}^{k}$, for $q\geq q^{\prime}$; (b) $A_{qq^{\prime}}^{<k}=A_{qq^{\prime}}^{k}$, for $q^{\prime}\geq q$, then

\[
A_{qq^{\prime}}^{>k}\left(\gamma\right)=\frac{\mathrm{i}^{q-q^{\prime}}}{\left(q-q^{\prime}\right)!}\left[\frac{\left(k+q\right)!\left(k-q^{\prime}\right)!}{\left(k+q^{\prime}\right)!\left(k-q\right)!}\right]^{\frac{1}{2}}\left\{ \tan\left[{\scriptstyle \frac{1}{2}}\left(\theta_{1}-\gamma\theta_{2}\right)\right]\right\} ^{q-q^{\prime}}\left\{ \cos\left[{\scriptstyle \frac{1}{2}}\left(\theta_{1}-\gamma\theta_{2}\right)\right]\right\} ^{-2-2k}\]

\begin{equation}
\times\:_{2}F_{1}\left(k+1+q,k+1-q^{\prime};1+q-q^{\prime};-\tan^{2}\left[{\scriptstyle \frac{1}{2}}\left(\theta_{1}-\gamma\theta_{2}\right)\right]\right),\label{eq:A06}\end{equation}

\begin{equation}
A_{qq^{\prime}}^{<k}\left(\gamma\right)=A_{q^{\prime}q}^{>k}\left(\gamma\right).\label{eq:A07}\end{equation}

\noindent{}Substituting $A$ in Eq. (\ref{eq:4}) we gather

\begin{equation}
\left(n,n^{\prime};\alpha,\beta,\gamma,\delta|\hat{x}_{1},\hat{x}_{2}\right)_{qq^{\prime}}^{k}=\left\{ \begin{array}{ll}
\left(n,n^{\prime};\alpha,\beta,\gamma,\delta|\hat{x}_{1},\hat{x}_{2}\right)_{qq^{\prime}}^{>k}, & \quad q\geq q^{\prime},\\
\left(n,n^{\prime};\alpha,\beta,\gamma,\delta|\hat{x}_{1},\hat{x}_{2}\right)_{qq^{\prime}}^{<k}, & \quad q\leq q^{\prime},\end{array}\right. \label{eq:3.6.19}\end{equation}

\begin{equation}
\left(n,n^{\prime};\alpha,\beta,\gamma,\delta|\hat{x}_{1},\hat{x}_{2}\right)_{qq^{\prime}}^{>k}=\mathrm{i}^{\alpha q+\delta q^{\prime}}(-1)^{2\left(nk+n^{\prime}q^{\prime}\right)}\beta^{q^{\prime}-q}\mathrm{e}^{\mathrm{i}\left(q\varphi_{2}-q^{\prime}\varphi_{1}\right)}A_{qq^{\prime}}^{>k}\left(\gamma\right),\label{eq:A09}\end{equation}

\begin{equation}
\left(n,n^{\prime};\alpha,\beta,\gamma,\delta|\hat{x}_{1},\hat{x}_{2}\right)_{qq^{\prime}}^{<k}=\mathrm{i}^{\left(\alpha-\delta\right)\left(q-q^{\prime}\right)}\mathrm{e}^{\mathrm{i}\left(\varphi_{1}+\varphi_{2}\right)\left(q-q^{\prime}\right)}\left(n,n^{\prime};\alpha,\beta,\gamma,\delta|\hat{x}_{1},\hat{x}_{2}\right)_{q^{\prime}q}^{>k}.\label{eq:A010}\end{equation}

\section{\label{C}Reduced matrix element on the basis of spherical harmonics}

Here a proof of Eq. (\ref{eq:ccred2}), exploiting RCGC technique, will be offered. The matrix element of $T_{q}^{k}\left(\hat{x}\right)$ on the basis of arbitrary functions $\psi_{m}^{l\dagger}(\hat{x})$ and $\psi_{\bar{m}}^{\bar{l}}(\hat{x})$ is written by

\begin{equation}
\langle lm\vert T_{q}^{k}\vert\bar{l}\bar{m}\rangle=[l\Vert T^{k}\Vert\bar{l}]\left[\begin{array}{ccc}
\bar{l} & k & l\\
\bar{m} & q & m\end{array}\right]={\displaystyle \int_{S^{2}}}\mathrm{d}\hat{x}\: \psi_{m}^{l\dagger}(\hat{x})T_{q}^{k}(\hat{x})\psi_{\bar{m}}^{\bar{l}}(\hat{x}).\label{eq:C1}\end{equation}

\noindent{}Direct adaptation of Eq. (\ref{eq:tr17}) points to

\begin{equation}
\langle lm\vert T_{q}^{k}\vert\bar{l}\bar{m}\rangle={\displaystyle \int_{S^{2}}}\mathrm{d}\hat{x}\:{\displaystyle \sum_{m^{\prime}q^{\prime}\bar{m}^{\prime}}}\psi_{m^{\prime}}^{l\dagger}(\hat{x}^{\prime})T_{q^{\prime}}^{k}(\hat{x}^{\prime})\psi_{\bar{m}^{\prime}}^{\bar{l}}(\hat{x}^{\prime})\overline{\eta_{mm^{\prime}}^{l}(\hat{x}^{\prime},\hat{x})}\eta_{qq^{\prime}}^{k}(\hat{x}^{\prime},\hat{x})\eta_{\bar{m}\bar{m}^{\prime}}^{\bar{l}}(\hat{x}^{\prime},\hat{x}).\label{eq:C2}\end{equation}

\noindent{}Transformed functions $\psi$ and operator $T$ depend on fixed coordinates $\hat{x}^{\prime}$. Consequently, they can be located in front of the integral. Recalling, that $\overline{\eta_{mm^{\prime}}^{l}}=(-1)^{m-m^{\prime}}\eta_{-m-m^{\prime}}^{l}$, and exploiting reduction rules for the Kronecker products $l\times k\rightarrow \bar{L}$, $\bar{L}\times\bar{l}\rightarrow L$ (see Eq. (\ref{eq:18})), we attain $\eta_{MM^{\prime}}^{L}(\hat{x}^{\prime},\hat{x})$. The integral of present spherical function is defined in Eq. (\ref{eq:S2c}), and in this case equals to $\mathcal{S}_{0M^{\prime}}^{L}(\hat{x}^{\prime};+)$ (see Eq. (\ref{eq:S2d})), since $L\in\mathbb{Z}^{+}$. Thus

\[
\langle lm\vert T_{q}^{k}\vert\bar{l}\bar{m}\rangle={\displaystyle \sum_{m^{\prime}q^{\prime}\bar{m}^{\prime}}}(-1)^{m-m^{\prime}}\psi_{m^{\prime}}^{l\dagger}(\hat{x}^{\prime})T_{q^{\prime}}^{k}(\hat{x}^{\prime})\mathcal{S}_{0M^{\prime}}^{L}(\hat{x}^{\prime};+)\psi_{\bar{m}^{\prime}}^{\bar{l}}(\hat{x}^{\prime})\]

\begin{equation}
\times\left[\begin{array}{ccc}
l & k & \bar{L}\\
-m & q & -\bar{m}\end{array}\right]\left[\begin{array}{ccc}
l & k & \bar{L}\\
-m^{\prime} & q^{\prime} & \bar{M}^{\prime}\end{array}\right]\left[\begin{array}{ccc}
\bar{L} & \bar{l} & L\\
-\bar{m} & \bar{m} & 0\end{array}\right]\left[\begin{array}{ccc}
\bar{L} & \bar{l} & L\\
\bar{M}^{\prime} & \bar{m}^{\prime} & M^{\prime}\end{array}\right].\label{eq:C3}\end{equation}

\noindent{}The matrix element does not depend on $\hat{x}^{\prime}$, thus one can choose any value. We select the most simple case $\hat{x}^{\prime}=\hat{0}\equiv (0,0)$. The application of orthogonality condition for Clebsh-Gordan coefficients, combining Eqs. (\ref{eq:C1}), (\ref{eq:C3}), leads to equalities $\bar{L}=\bar{l}$, $L=0$; thus $\mathcal{S}_{00}^{0}(\hat{0};+)=4\pi$ and

\begin{equation}
[l\Vert T^{k}\Vert\bar{l}]=\frac{4\pi}{2l+1}{\displaystyle \sum_{m^{\prime}q^{\prime}\bar{m}^{\prime}}}\psi_{m^{\prime}}^{l\dagger}(\hat{0})T_{q^{\prime}}^{k}(\hat{0})\psi_{\bar{m}^{\prime}}^{\bar{l}}(\hat{0})\left[\begin{array}{ccc}
\bar{l} & k & l\\
\bar{m}^{\prime} & q^{\prime} & m^{\prime}\end{array}\right].\label{eq:C4}\end{equation}

\noindent{}It is noticeable, obtained expression, by applying technique of coordinate transformations, coincides with Eq. (41) in \cite{Rudzkas2}, if reducing given Kronecker products $l\times k\rightarrow L^{\prime}$, $L^{\prime}\times \bar{l}\rightarrow L^{\prime\prime}$. This would signify $L^{\prime}=\bar{l}$ and $L^{\prime\prime}=0$. At this step we turn to special case of eigenfunctions $\psi_{m}^{l}(\hat{x})=Y_{m}^{l}(\hat{x})$, what implies equality $Y_{m}^{l}(\hat{0})=\delta_{m0}\mathrm{i}^{l}\sqrt{(2l+1)/4\pi}$. Hence Eq. (\ref{eq:C4}) becomes equal to expression, presented in Eq. (\ref{eq:ccred2}). One should be also reminded, that in general, in Eq. (\ref{eq:C4}) the arguments $\hat{0}$ can be replaced by any values $\hat{x}$.

%\begin{acknowledgements}
%\end{acknowledgements}



\end{document}